\documentclass{WileyMSP-template}

\usepackage{siunitx}
\usepackage{graphicx}
\usepackage{amsmath}
\usepackage{amssymb}

\usepackage[numbers,sort&compress]{natbib}

\begin{document}

\pagestyle{fancy}
\rhead{\includegraphics[width=2.5cm]{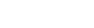}}

\title{Structural Changes and Transport Properties of {YBa}$_2${Cu}$_3${O}$_7$ \\ Locally Modified by a He$^+$ Focused Ion Beam}

\maketitle


\author{Ross~Carter$^{1,*}$}%
\author{Robin~Hutt$^{2,*}$}%
\author{Paul~Zimmermann$^1$}%
\author{Ainur~Abukaev$^1$}%
\author{Jan~Ullmann$^2$}%
\author{Simon~Koch$^2$}%
\author{Christoph Schmid$^2$}%
\author{Manfred~Burghammer$^3$}%
\author{Reinhold~Kleiner$^2$}%
\author{Dieter~Koelle$^{2,4,5}$}%
\author{Edward~Goldobin$^{2, \dagger}$}%
\author{Ivan~A.~Zaluzhnyy$^{1, \ddagger}$}%

\begin{affiliations}
$^1$ Institute of Applied Physics, University of T\"ubingen, Auf der Morgenstelle 10, 72076 T\"ubingen, Germany

$^2$ Institute of Physics, University of T\"ubingen, Auf der Morgenstelle 14, 72076 T\"ubingen, Germany

$^3$ European Synchrotron Radiation Facility, 71 Avenue des Martyrs, 38000 Grenoble, France

$^4$ Center for Quantum Science, University of T\"ubingen, Auf der Morgenstelle 14, 72076 T\"ubingen, Germany

$^5$ Center for Light-Matter Interaction, Sensors \& Analytics LISA+, University of T\"ubingen, Auf der Morgenstelle 15, 72076 T\"ubingen, Germany

$^*$ These authors contributed equally

$^\dagger$ gold@uni-tuebingen.de

$^\ddagger$ ivan.zaluzhnyy@uni-tuebingen.de

\end{affiliations}


\begin{abstract}
\justifying
Irradiation of a material with ions can cause various defects that can lead to structural phase transitions and the modification of the material's properties. 
Here we study the irradiation of the epitaxyally grown thin films of the high-temperature superconductor YBa$_2$Cu$_3$O$_{7}$ with $30$~keV He$^{+}$ ions which leads to the expansion of the crystal lattice, decrease of the critical temperature $T_c$ and eventually transition to an insulator. Fabrication of such insulating regions with a focused He-Ion beam with a spot size of $\sim10$~nm is a powerful technique for fabrication of superconducting nano-devices. Using low-temperature resistivity measurements, diffraction with a nanofocused X-ray beam and atomic force microscopy, we investigated how the structure and the electric transport properties of YBa$_2$Cu$_3$O$_{7}$ depend on the irradiation dose in a range 10--100~ions/nm$^2$ and on the lateral size of the irradiated area in a range 30--5000~nm.

\end{abstract}


\justifying

	\section{Introduction}
\label{sec:intro}
The ability to change the electrical properties of a material on the nanoscale is a crucial technological process in superconducting electronics.
The irradiation of superconducting thin films with a focused He-ion beam (He-FIB) was established recently as a new nano-fabrication technique, which was subsequently applied to various superconducting materials \cite{Cybart:2015:He-FIB.JJs}.
For example, using Nb films one can fine-cut constriction-type Josephson junctions (cJJs) and create superconducting quantum interference devices (SQUIDs) based on them \cite{Weber:2025:SQUID-on-lever}.
Josephson junctions with a barrier layer (bJJs) formed by He-FIB irradiation were created based on MgB$_2$ films \cite{Kasaei:2018:He-FIB:MgB2-JJ, Kasaei:2019:bJJA.LowSpread, Yin:2024:He-FIB:MgB2-JJ}, Co-doped BaFe$_2$As$_2$ films \cite{Chen:2024:bJJs(Co-BaFeS)} and NbTiN films \cite{Ruhtinas:2025:He-FIB:NbTiN:bJJs}.

Especially promising is the use of He-FIB to modify the properties of cuprate superconductors, with YBa$_2$Cu$_3$O$_{7-\delta}$ (YBCO) being a particularly prominent example. 
He-FIB was utilized to create artificial pinning nano-arrays with a lattice spacing as low as \qty{20}{\nm} (matching field of \qty{6}{\tesla})\cite{Karrer:2024:He-FIB:PinArr20nm}, which allowed for the demonstration of new physical effects, such as the formation of an unusual Bose glass of vortices \cite{Aichner:2023:MagRes(ang), Backmeister:2022:He-FIB:YBCO:BoseGlass}.
Several groups have demonstrated an excellent functionality of bJJs \cite{Cybart:2015:He-FIB.JJs, Cho:2018:He-FIB:bJJs(w), Mueller:2019:He-FIB:JJ&SQUID, Couedo:2020:He-FIB:YBCO-JJs, Chen:2022:YBCO-bJJs} and SQUIDs \cite{Mueller:2019:He-FIB:JJ&SQUID, Cho:2015:He-FIB.SQUID} fabricated by He-FIB, and investigated THz properties and Shapiro steps for bJJs \cite{Proepper:2024:bJJ.THz} and bJJ arrays \cite{Proepper:2024:He-FIB:YBCO:JJA}.
Recently more advanced devices, for example Josephson diodes, have been demonstrated with record figures of merit \cite{Schmid:2025:He-FIB:YBCO-JJD}, as well as constriction-type Josephson junctions (cJJs) with a constriction width comparable with the He-FIB spot size $\sim$\qty{10}{\nm} \cite{Schmid:He-FIB:YBCO:cJJ&SQUID}.

Despite the substantial use of the He-FIB in superconducting nano-electronics, the structural changes caused by the ion irradiation at nanoscale, and their relation to the electronic properties, remain obscure.
It has been established that the He$^+$ ions create long-living defects in the YBCO crystal lattice, with a radiation-dose-dependent lifetime \cite{Karrer:2024:bJJ(t)}. 
If we consider only single lines ``drawn'' by He-FIB, it was assumed up to now that lower line-dose irradiation ($\lesssim200$~ions/nm) displaces some of the chain O-atoms, thus changing the effective doping of YBCO and reducing the critical temperature $T_c$ \cite{Valles1989, Gupta1992, Menushenkov1995a, Navacerrada2000, Nicholls2022, Gray2022}.
Moderate line-dose irradiation ($\sim$ 400--700~ions/nm) suppresses $T_c$ down to zero.
Consequently, since the He-FIB line has a width of only few nanometres, one can observe Josephson tunnelling through such a barrier, \textit{i.e.}, obtain a bJJ.
For still higher line-dose values, the Josephson critical current density decreases and barrier resistance increases exponentially with the dose \cite{Mueller:2019:He-FIB:JJ&SQUID}.
For line-doses $\gtrsim1500$~ions/nm, the number of defects within the irradiated line becomes so large that the YBCO crystal turns into an amorphous material. 
This effect can be clearly seen by scanning electron transmission microscopy (STEM) on samples irradiated by ions, electrons and neutrons \cite{Karrer:2024:He-FIB:PinArr20nm,Mueller:2019:He-FIB:JJ&SQUID,Schmid:He-FIB:YBCO:cJJ&SQUID,Lee2016, Suvorova2014, Kirk1999, Hutt:AmorDose}.
In addition to the formation of point defects, the unit cell parameters $a$, $b$ and $c$ of the orthorhombic YBCO lattice also change on average upon irradiation \cite{Nicholls2022}.

Moreover, the structural changes in YBCO seem to depend on the size of the irradiated area \cite{Zaluzhnyy:2024:YBCO:He-FIB.StructChg}.
This does not allow us to directly apply the broad range of results obtained on large-area YBCO films irradiated with an unfocused ion beam \cite{Valles1989, Gupta1992, Terai1991, Zhao1991, Meyer1989, Navacerrada2000, Arias2003, Nicholls2022} to the nanopatterns fabricated by He-FIB \cite{Mletschnig2019,Li2024}.
Therefore, the studies of locally modified YBCO with sufficient spatial resolution are imperative to understand the correlation between crystal structure and superconducting properties.

In this work, we shed light on structural changes of YBCO upon He-FIB irradiation, focusing on dose-values below the threshold of amorphisation.
We report on the investigation of irradiated areas using different techniques: (a) local spatially resolved X-ray diffraction (XRD) studies with an X-ray spot size $\sim$\qtyproduct{100 x 100}{\square\nm}, (b) resistance $R$ vs. temperature $T$ measurements, and (c) atomic force microscopy (AFM) scans.
YBCO areas of varying length, ranging from $30$ to $5000$~nm, were irradiated with doses from 10 to 100~ions/nm$^2$.
We studied the dependence of both the in-plane and out-of-plane lattice parameters of YBCO on the irradiation dose and irradiation area, correlating the structural changes with the transport measurements performed on the same YBCO chip.

\section{Results and discussion}

\subsection{Samples}
\label{sec:samples}

\begin{figure}[]
	\begin{center}
		\includegraphics[width=0.5\linewidth]{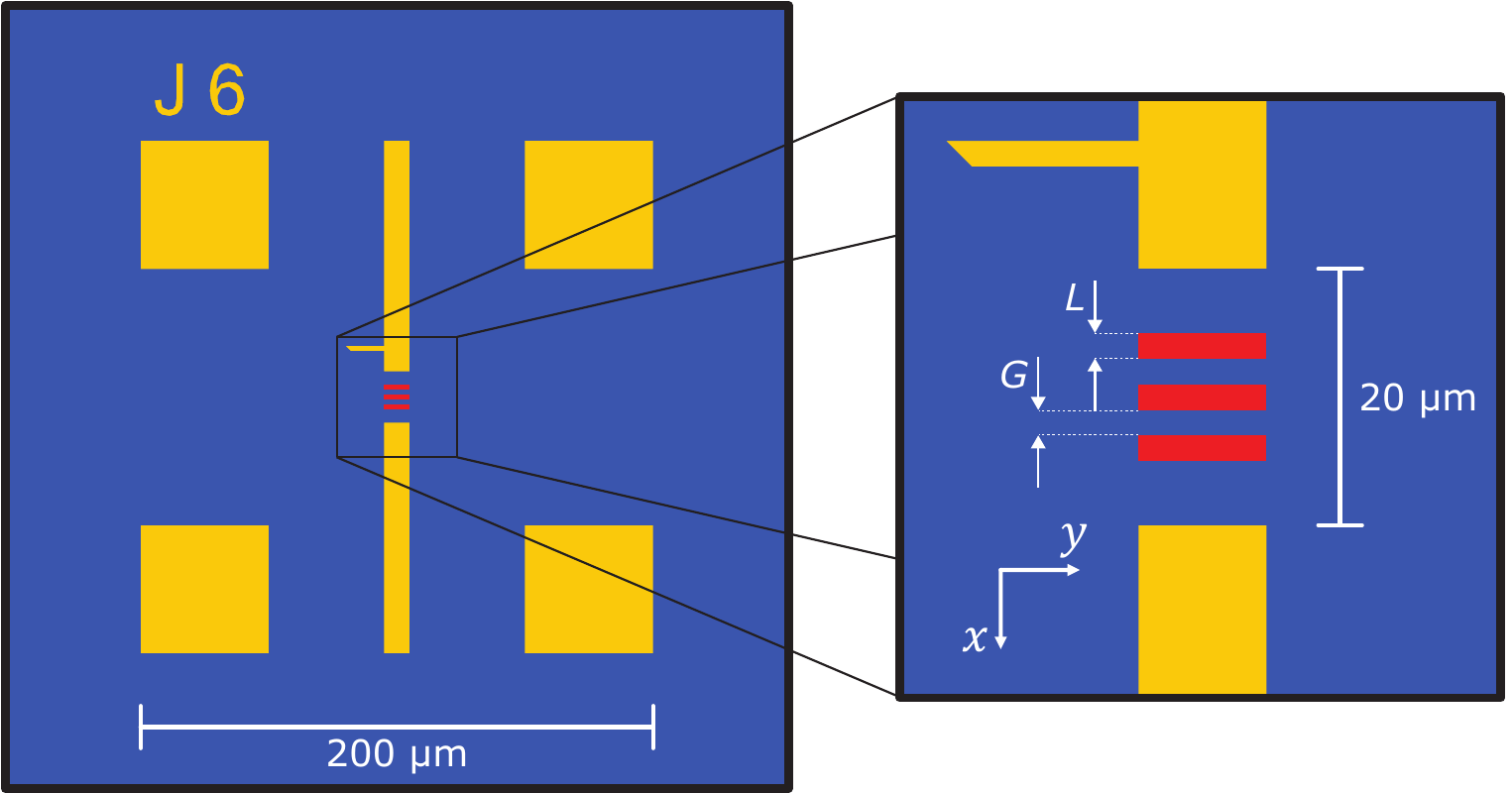}
	\end{center}
	\caption{%
		An example of one XRD block.
		On top of pristine YBCO (blue) there are Au (gold) alignment markers.
		The irradiation pattern for this XRD block consists of three stripes (red) of the size \qtyproduct{10 x 2}{\square\micro\metre} and spaced by \qty{2}{\um}. The dose values $D=10,\,20,\,30$~ions/nm$^2$ were used for this particular block.
		The inset shows a magnification of the irradiated area.
	}
	\label{Fig:XRD.block}
\end{figure}

\begin{figure}[]
	\begin{center}
		\includegraphics[width=0.5\linewidth]{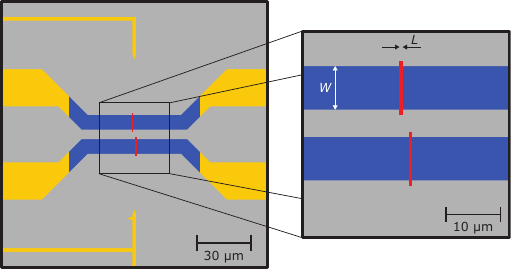}
	\end{center}
	\caption{%
		The geometry of micro bridges. After lithography and etching down to the LSAT substrate (light gray) we are left with the micro bridges consisting of a YBCO-Au bilayer of width $W$ (here $W$=\qty{8}{\um} and length \qty{50}{\um} connected to the contact pads.
		In the central part of each micro bridge, the Au layer is then removed so that one sees pristine YBCO (blue). These regions are used for irradiation by He-FIB. The irradiated stripes (red) have a length $L$ (here $L$=\qty{60}{\nm}) in the transport direction and exceed the bridge width $W$ to avoid shunting through the edges.
	}
	\label{Fig:Transp.sketch}
\end{figure}

We prepared epitaxially-grown $c$-axis oriented YBCO on (100)-oriented single crystal (LaAlO$_3$)$_3$(Sr$_2$AlTaO$_6$)$_7$ (LSAT) substrates (Figure S1-S2 in the Supporting Information (SI)).
The \qtyproduct{10 x 10}{\square\mm} chip was divided into two sections: a transport section, with devices to measure the resistivity of irradiated areas; and an XRD section for the spatially-resolved structural studies (see Figure~S3 in the SI).
The XRD section of the chip initially consisted of one vast YBCO area, connected to ground for charge sink during FIB irradiation, overlaid with Au. This area was then patterned (simultaneously with the transport part, see below) into an array of \qtyproduct{200 x 200}{\square\um} blocks (Figure~\ref{Fig:XRD.block}) by lithographically removing sections of Au, leaving only the alignment marks.
Thereafter, using He-FIB with ion energy 30~keV, each block was irradiated to create a series of stripes (rectangles $L\times$\qty{10}{\um}) of length $L$ interspaced with non-irradiated (pristine) YBCO spacings of length $G$. Each subsequent, irradiated stripe had an increasing dose $D$ in the range $10\ldots100$~ions/nm$^2$.
The values of $L$ and $G$ were different between the blocks starting from $L=G=500$~\unit{nm} so that one has ten $D$-values in the above range within one single block.
We also irradiated stripes with $L=G=1000,2000,5000$~nm, in which case one needs several XRD blocks to cover the whole $D$-range specified above.

The transport section of the chip consists of two identically structured \qtyproduct{5 x 5}{\square\mm} sub-chips (see Figure~S3 in the SI). 
Within each sub-chip,
we lithographically created a set of microbridges, with width $W=2$ or 8~\unit{\um} and length 50 or 100~\unit{\um}, connected to contact pads for 4-point measurements (Figure~\ref{Fig:Transp.sketch}).
We then irradiated rectangular areas across the microbridges using He-FIB, creating a set of structures with parameters summarized in Table~\ref{Tab:Transp.params}.

\begin{table}[!h]
	\begin{center}
	\begin{tabular}{ccc}
		$W$\,[\unit{\um}] & $L$\,[nm] & $D$\,[ions/nm$^2$]\\
		\hline
		2 & 30 & 10, 20 \ldots100\\
		8 & 150 & 10, 20 \ldots80\\
		8 & 300 & 10, 20 \ldots60\\
		8 & 420 & 10, 20 \ldots60\\
		8 & 600 & 10, 20 \ldots60\\
		\hline
	\end{tabular}
	\end{center}
	\caption{
		The parameters of the irradiated areas in the transport part of the chip. $W$ is the width of the microbridge, $L$ is the length of the irradiated rectangle in the transport direction.}
	\label{Tab:Transp.params}
\end{table}

\subsection{Nanofocused X-ray diffraction}
\label{Sec:LocalXRD}

\begin{figure}
	\begin{center}
		\includegraphics[width=0.5\linewidth]{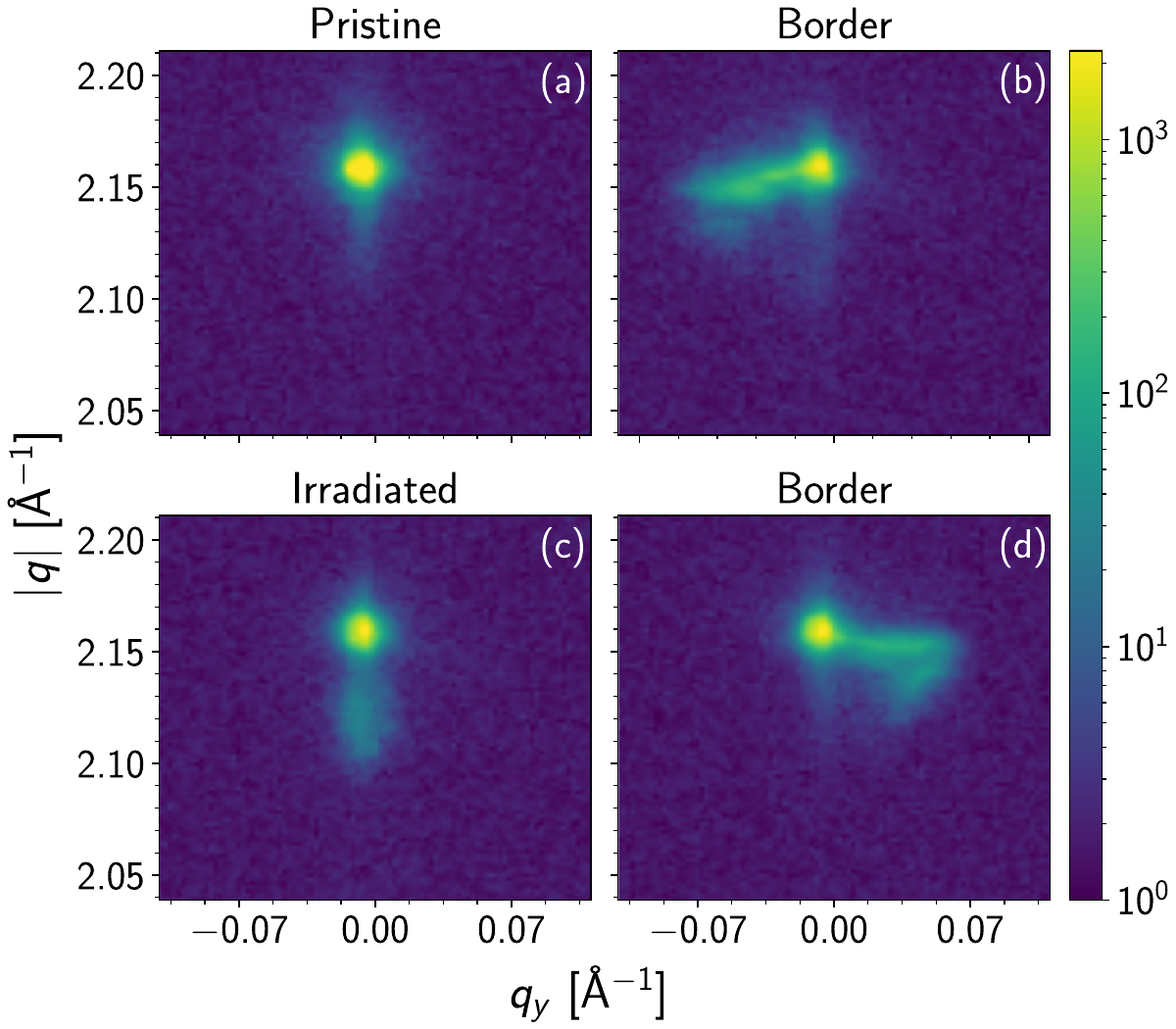}
	\end{center}
	\caption{%
		Scattered intensity $I(|q|,q_y)$ of the 004 diffraction peak of YBCO at different positions across an $L = 2000$~nm stripe which had been irradiated with $D = 25$~ions/nm$^2$.
		(a) An entirely pristine region of YBCO.
		(b) A transition region from pristine to irradiated YBCO, measured on the spacing-stripe boundary.
		(c) An entirely irradiated region.
		(d) A transition region from irradiated to pristine YBCO, measured on the opposite stripe-spacing boundary.
	}
	\label{fig:DetSpots}
\end{figure}

\begin{figure}
	\begin{center}
		\includegraphics[width=\linewidth]{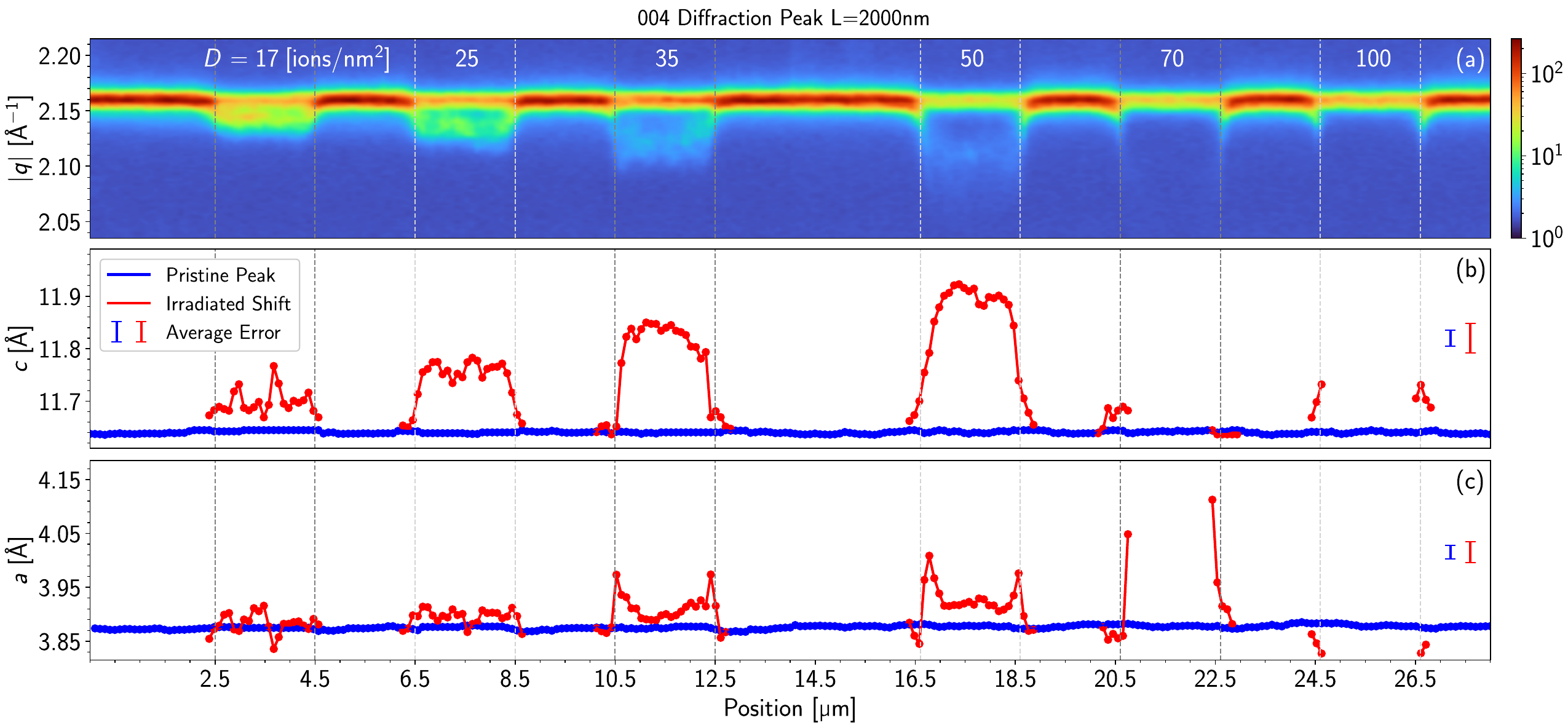}
	\end{center}
	\caption{
		(a) $|q|$-profile of the 004 peak of YBCO (integrated over $q_y$) at different positions across stripes of $L$=\qty{2000}{\nm} with increasing dose from left to right. The dashed vertical lines denote the boundaries of irradiated regions.
		(b-c) Swelling of the (b) out-of-plane lattice parameter, $c$ and (c) in-plane lattice parameter, $a$ across YBCO film with irradiated stripes of length $L$=\qty{2000}{\nm}.
	}
	\label{fig:ac(D).2000nm}
\end{figure}

We focused X-rays to the size of approximately \qtyproduct{100 x 100}{\square\nm} (FWHM) to record the scattering signal exclusively from a small area illuminated by the X-ray beam.
We selected two diffraction peaks of YBCO to measure at each spatial position, namely, $004$ and $104$ (Figure S2 in the SI), to obtain the out-of-plane lattice parameter, $c$, and the in-plane lattice parameter, $a$.
The specific choice of the reflections was done based on the following criteria: the peak should be above the sample's horizon, it should not overlap with the reflections from LSAT or Au, and it should have enough intensity for effective analysis.

For the analysis of the YBCO lattice parameters, we aligned the sample to the Bragg condition by choosing an appropriate orientation of the substrate, determined by the angle $\omega$, and scanned the He-FIB fabricated stripes with a nanofocused X-ray beam. We averaged the diffraction patterns measured at the same $x$-position across the stripes over $y$ (see Figure S2b in the SI).
Subsequently, we re-sampled the averaged diffraction patterns from detector coordinates to $q$-space using the \texttt{pygid} package \cite{Abukaev2025} and custom, author-developed scripts.
The examples of $I(|q|,q_y)$ in the selected region of interest (ROI) around the 004 diffraction peak, where $|q|=\sqrt{q_x^2+q_y^2+q_z^2}$, are shown in Figure~\ref{fig:DetSpots}. Each panel shows how the reflection changes depending on where the X-ray beam illuminates, relative to an irradiated stripe.

We then employed 2-dimensional Gaussian fitting on each 004 and 104 diffraction peak to extract their parameters, most importantly its position $q=|\mathbf{q}|$.
Where appropriate, e.g. the diffraction pattern in Figure~\ref{fig:DetSpots}c, we treat the pattern as two separate peaks, one pristine peak characterised by a $q$-position consistent with pristine YBCO, and one diffuse peak at smaller $q$ values. In these cases, two distinct Gaussian fits were used to obtain the $q$-positions of both peaks.

\begin{figure*}
	\begin{center}
		\includegraphics[width=\linewidth]{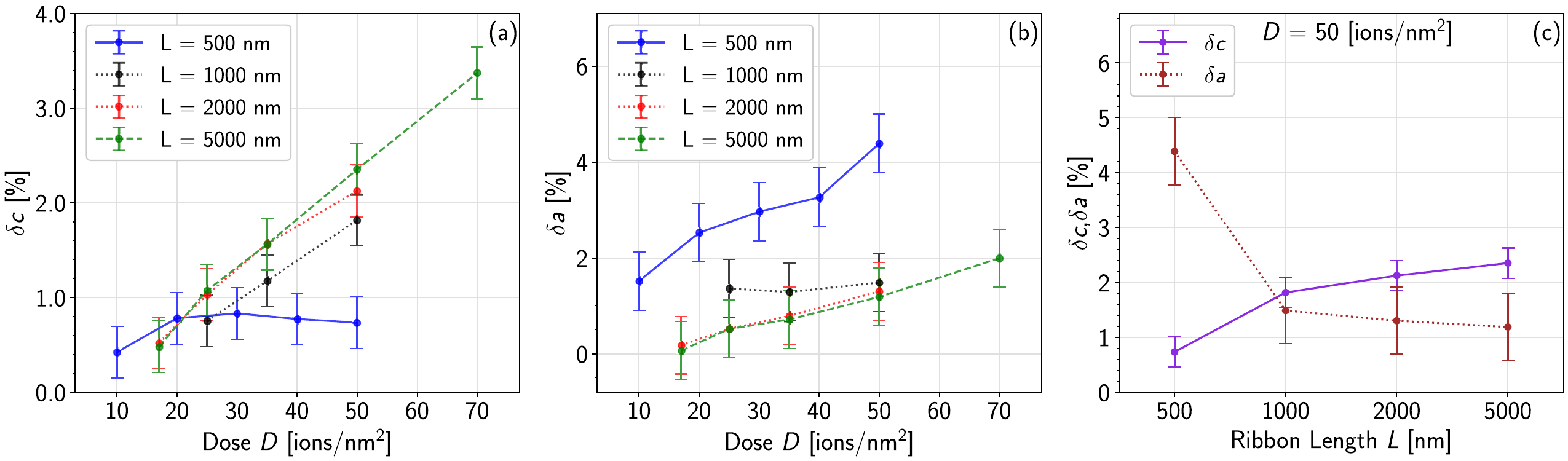}
	\end{center}
	\caption{%
		(a-b) Relative change in lattice constants $\delta c$ (a) and $\delta a$ (b) for different doses $D\le D_{dis}^{(00l)}$, the critical-disorder dose in the $(00l)$ direction, measured at each $L$.
		(c) A comparison of $\delta c$ and $\delta a$ for different stripe lengths $L$, at a constant dose $D=50$~ions/nm$^2$.
	}
	\label{fig:ac(D)}
\end{figure*}

Figure~\ref{fig:ac(D).2000nm}a shows the $q$-profile of the 004 peak of YBCO for stripes with $L$=\qty{2000}{\nm}. An immediate observation is that the pristine peak remains visible for all $D$. However, for $D>50$~ions/nm$^2$ its intensity decreases, which is likely related to the partial amorphisation of the YBCO crystal lattice.
We attribute the persistence of the weaker, but still relatively sharp, pristine peak observable for $D>50$~ions/nm$^2$ to nanoscale grains (fragments) of the YBCO crystal lattice that retain their own internal structure, even inside the irradiated stripes.
This is consistent with previous research \cite{Zaluzhnyy:2024:YBCO:He-FIB.StructChg}, where it was observed that some small fraction of pristine YBCO remained in irradiated areas and stripes when irradiating with an FIB. 
This effect is possibly owing to the local nature of irradiation when using a FIB, as the irradiated material cannot freely expand due to contact with neighbouring (non-irradiated) regions, so the pristine crystal structure of YBCO is stabilized close to the substrate.
The pristine peak persists up to the highest irradiation dose $D=100$~ions/nm$^2$, and only disappears at a much higher amorphisation dose $D_c$.
In the STEM studies \cite{Karrer:2024:He-FIB:PinArr20nm,Mueller:2019:He-FIB:JJ&SQUID,Schmid:He-FIB:YBCO:cJJ&SQUID}, it was observed that when $D>D_c$, the crystal amorphises in the irradiated region; here we observe similar effects, albeit less pronounced due to the smaller doses used.

The appearance of a diffuse peak at slightly smaller $q$-values than those for the pristine peak indicates that some fraction of the YBCO lattice expands and loses its perfect crystal order due to the formation of structural defects.
This peak undergoes a greater change in $q$ with increasing irradiation dose $D$, reaching a maximum change at a critical value, $D_{dis}$.
At doses $D>D_{dis}$, the disorder in the YBCO crystal lattice is so high that the diffuse peak becomes indistinguishable from the background, and the signal-to-noise is so small that reliably fitting the peak is not possible.
We therefore define the critical-disorder dose, $D_{dis}$, as the maximum dose for which we still observe a diffuse peak; notably this dose is much smaller than the previously described amorphisation dose, $D_c$.
One exception to this rule is the boundary between pristine and irradiated regions, see Figure~\ref{fig:DetSpots}b,d. Here, the behaviour of the diffuse peak more closely resembles the previously discussed behaviour of the pristine peak, whereby it remains even at the highest irradiation dose $D=100$~ions/nm$^2$. This is likely because the neighbouring pristine region anchors the irradiated YBCO, which prevents complete disorder and thus the erosion of the diffuse Bragg peak.

The $q$-position of the 004 peak allows us to directly calculate the lattice constant $c=4\cdot 2\pi/q_{004}$, presented in Figure~\ref{fig:ac(D).2000nm}b.
The lattice in this example expands by approximately 0.2~\AA~in the $c$-direction at $D = 50$~ions/nm$^2$, in agreement with previous works with unfocused \cite{Terai1991, Nicholls2022} and focused \cite{Zaluzhnyy:2024:YBCO:He-FIB.StructChg} ion beams.
Subsequently, we investigated the 104 peak utilising the same methods as above (see Figure~S5 in the SI). Using the $q$-position of the 104 peak in conjuncture with the aforecalculated out-of-plane lattice constant $c$, we determined the in-plane lattice constant
\begin{equation}
	a = \frac{1}{\sqrt{\Big(\frac{q_{104}}{2\pi}\Big)^2-\Big(\frac{4}{c}\Big)^2}},
\end{equation}
which is shown in Figure~\ref{fig:ac(D).2000nm}c. The value of $a$ increases by $\approx0.1$~\AA ~at the same dose.
Such an increase of the in-plane lattice parameter for small irradiated areas is drastically different from the only small changes of the average in-plane lattice parameter observed for unfocused ion irradiation \cite{Zaluzhnyy:2024:YBCO:He-FIB.StructChg}.
We explain this unexpected expansion of the in-plane lattice parameter by the bending discussed in section \ref{Sec:afm}.
This difference in the behaviour of the in-plane lattice parameter for focused and unfocused ion irradiation indicates that the strain at the boundary between the irradiated and pristine YBCO can alter the structural transition in YBCO and modify its properties in dependence of the size and shape of the irradiated area.

\begin{table}
		\begin{center}
	\begin{tabular}{ccc}
		Peak \, & $L$\,[nm] & $D_{dis}$\,[ions/nm$^2$]\\
		\hline
		007 & 500 & 50\\
		004 & 1000 & 50\\
		004 & 2000 & 50\\
		004 & 5000 & 70\\
		\hline
		104 & 500 & 60\\
		104 & 1000 & 70\\
		104 & 2000 & 70\\
		104 & 5000 & 70\\
		\hline
	\end{tabular}
		\end{center}
	\caption{Summary of each investigated diffraction peak and the corresponding critical-disorder dose $D_{dis}$.
	}
	\label{Tab:Disorder_limit}
\end{table}

The same analysis was also employed for stripes of $L = 500, 1000$~and $5000$~nm (see Figs.~S5--S10  in the SI). 
Figure~\ref{fig:ac(D)} summarises the observed changes in the lattice parameters as a function of dose $D$ and length $L$ (see also Figs.~S11--S13  in the SI).
Here, $\delta c=\Delta c/c$ and $\delta a =\Delta a/a$ represent the absolute change in lattice constants $c$ and $a$ respectively, averaged across each irradiated stripe (along $x$-axis in Figure~S2b  in the SI, excluding stripe-spacing boundaries). 
At $L = 500$~nm, $\delta a$ is approximately $4.4\%$ at $D=50$~ions/nm$^2$, significantly greater than observed in unfocused (broad beam) irradiation studies \cite{Zaluzhnyy:2024:YBCO:He-FIB.StructChg} or when changing the oxygen deficiency in YBa$_2$Cu$_3$O$_{7-\delta}$ \cite{Jorgensen1990,Cava1990}.
Between $L = 500$ and 1000~nm there is a steep decrease in $\delta a$, before a more gradual decline as $L$ increases; juxtaposed to this, $\delta c$ observes a sharp increase in the same $L$ range, before increasing more gradually at larger $L$.

Another key finding is that the value of the critical-disorder dose, $D_{dis}$, obeys two key trends. Firstly, it tends to be slightly higher for larger values of $L$; secondly, it is always smaller for the 004 Bragg peak than for the 104 Bragg peak (Table~\ref{Tab:Disorder_limit}). The latter trend suggests that the defects induced by the He-FIB irradiation destroy the crystal order in the out-of-plane direction ($c$-direction) first.

One caveat to the methods presented in this chapter is that we require the out-of-plane lattice constant $c$ to calculate the in-plane lattice constant $a$, meaning that we can only reliably measure $a$ up to the critical-disorder dose in the $(00l)$ direction, $D_{dis}^{(00l)}$, despite the critical-disorder dose being larger in the $(h0l)$ direction.
Furthermore, the YBCO film twinning \cite{Khoshnevisan2002} along $\langle 110 \rangle$ directions did not allow us to distinguish the $a$ and $b$ orthorhombic lattice parameters.

\subsection{Atomic force microscopy}
\label{Sec:afm}

\begin{figure}
	\begin{center}
		\includegraphics[width=0.5\linewidth]{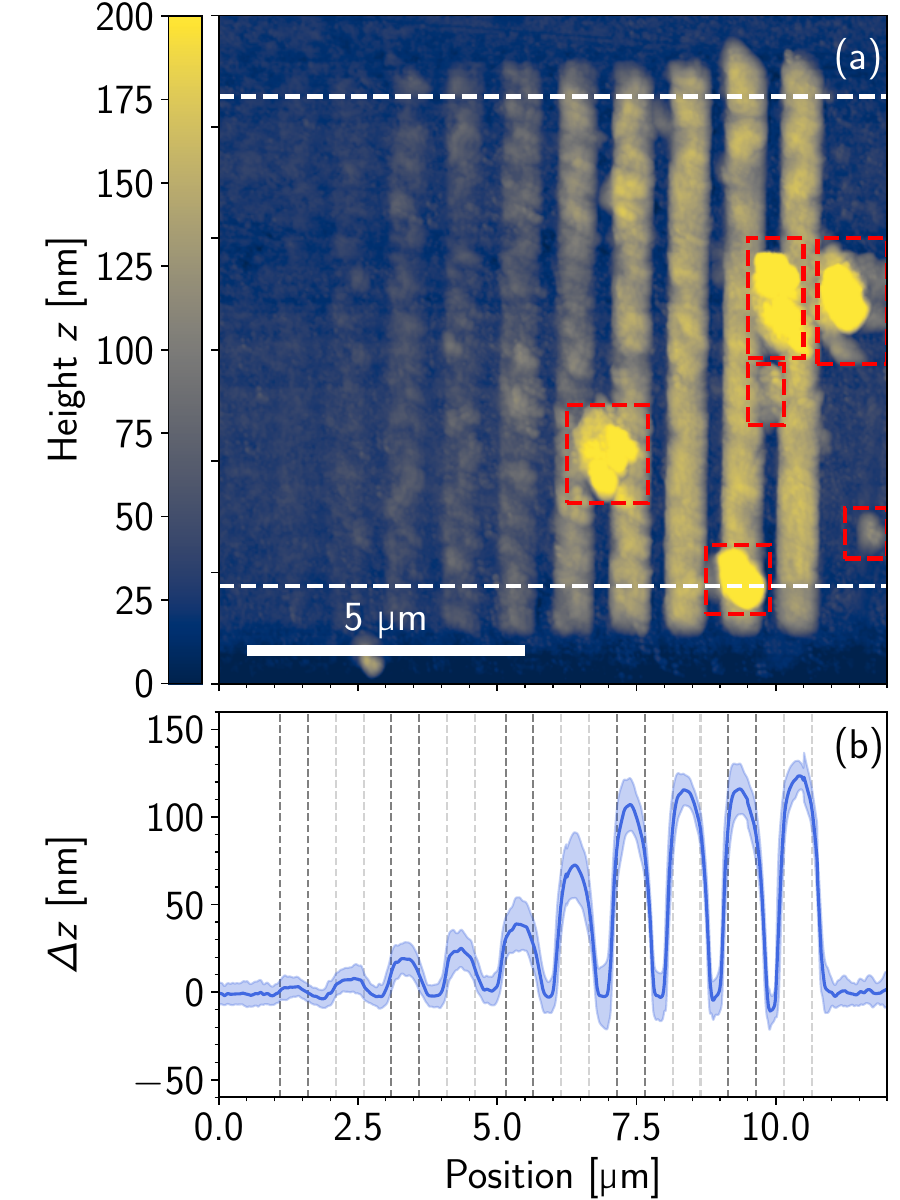}
	\end{center}
	\caption{%
		(a) AFM image (height $z$) measured across the 10 stripes of length $L=500$~nm irradiated with doses $D=10, 20, \ldots, 100$~ions/nm$^2$.
		The dashed, white lines represent the limits of the average profile in (b), the red boxes represent masked regions used to eliminate artefacts.
		(b) Change of the height $\Delta z$ relative to the pristine YBCO region. The shaded area represents the roughness of the film. The vertical dash lines indicate the borders of irradiated stripes.
	}
	\label{fig:YBCO_AFM}
\end{figure}

Figure~\ref{fig:YBCO_AFM}a shows the height $z$ of ten iradiated stripes measured by the AFM.
Here, we observe an increase in height of our film in irradiated regions up to $\Delta z\approx$\qty{130}{\nm} for the dose $D\approx 80$~ions/nm$^2$, after which $\Delta z$ saturates (Figure~\ref{fig:YBCO_AFM}b). 
This \emph{bending} up is far larger than the \emph{swelling} --- predicted changes due to an increase in $c$, which would have been on the order of a few nm.
This strongly indicates up-bending of both YBCO film and the LSAT substrate, probably caused by the formation of He ion bubbles deep inside LSAT.

The amount of bending appears to be the same for all $L\ge500$~nm for the doses $D<50$~ions/nm$^2$.
For higher doses ($D\ge50$~ions/nm$^2$), the change in height $\Delta z$ due to bending is maximal at $L = 500$~nm, and is slightly less pronounced for both smaller and larger $L$ (see Figure~S16  in the SI).

At the dose of $D=50$~ions/nm$^2$, bending was not observed at $L$=\qty{30}{\nm}, the narrowest stripe that was measured with AFM (Figure~S16  in the SI). This indicates that if $L$ is small enough then the restrictive boundaries of non-irradiated YBCO+LSAT prevent any such bending.
Additionally, this correlates with STEM images taken on the same $L$=\qty{30}{\nm} stripe, where no significant change in the stripe was observed \cite{Hutt:AmorDose}.

\subsection{Transport measurements}
\label{Sec:transport}

\begin{figure}
	\begin{center}
		\includegraphics[width=0.5\linewidth]{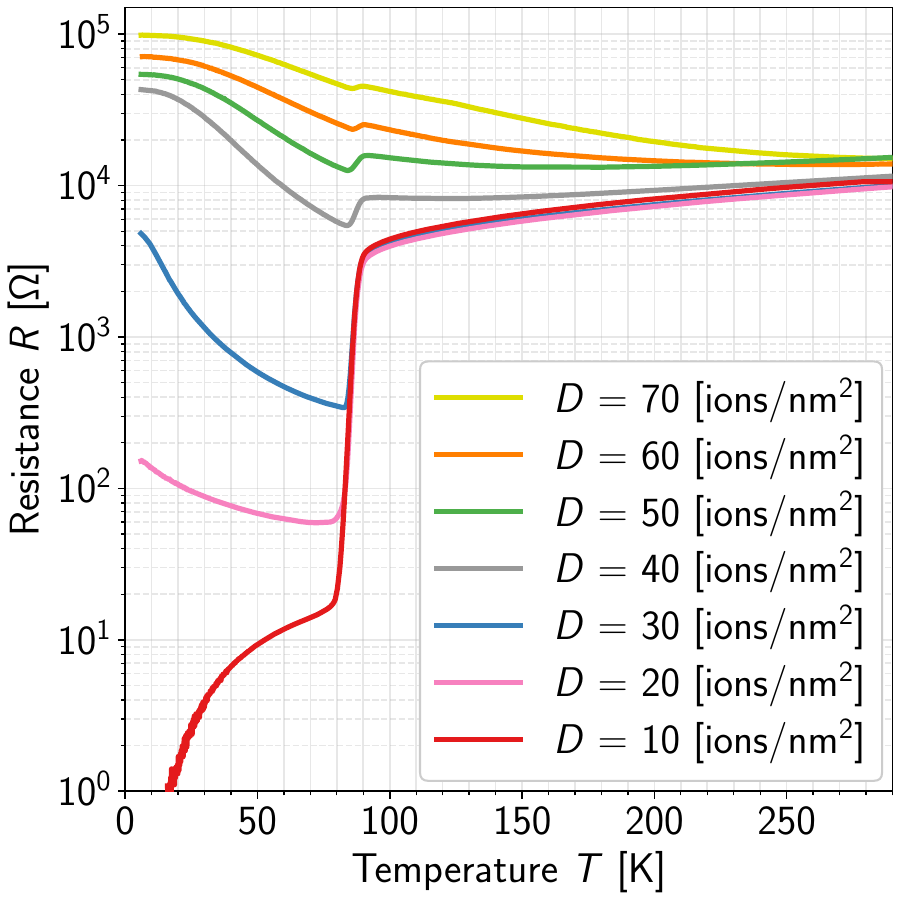}
	\end{center}
	\caption{%
		$R(T)$ curves ($I_b=1~\mu$A) of YBCO microbridges irradiated with different values of $D$ in the area $L \times W =$ \qty{30}{\nm} $\times$ \qty{2}{\um}.
	}
	\label{Fig:R(T)2x30}
\end{figure}

Figure~\ref{Fig:R(T)2x30} shows the $R(T)$ curves measured on fabricated microbridges (Figure~\ref{Fig:Transp.sketch}). One can see that below $T_\mathrm{c}$ the resistance of irradiated areas grows quite rapidly over several orders of magnitude with $D$ (which is correlated with structural changes, see, for example, Figure~\ref{fig:ac(D).2000nm}a) at a given $T$. 
This is compatible with a transport via variable-range hopping \cite{Lesueur:1993:YBCO:SIT(He-ions)}. 
For low values of $D$, \textit{e.g.}, $D=30$~ions/nm$^2$, the temperature plays an important role in the transport properties, such that $R$ changes one order of magnitude between $T=10$ and 80~K, however, the effect of $T$ is much weaker for large $D$.

We have also performed similar $R(T)$ measurements for bridges housing irradiated areas of different sizes, see Table~\ref{Tab:Transp.params}. The $R(T)$ dependences show similar behaviour; however, the resistances proved to be much larger than expected from the scaling with $L/W$. Many $R(T)$ curves for $D>40$~ions/nm$^2$ could not be measured reliably due to saturation limits of the setup.

\begin{figure}
	\begin{center}
		\includegraphics[width=0.5\linewidth]{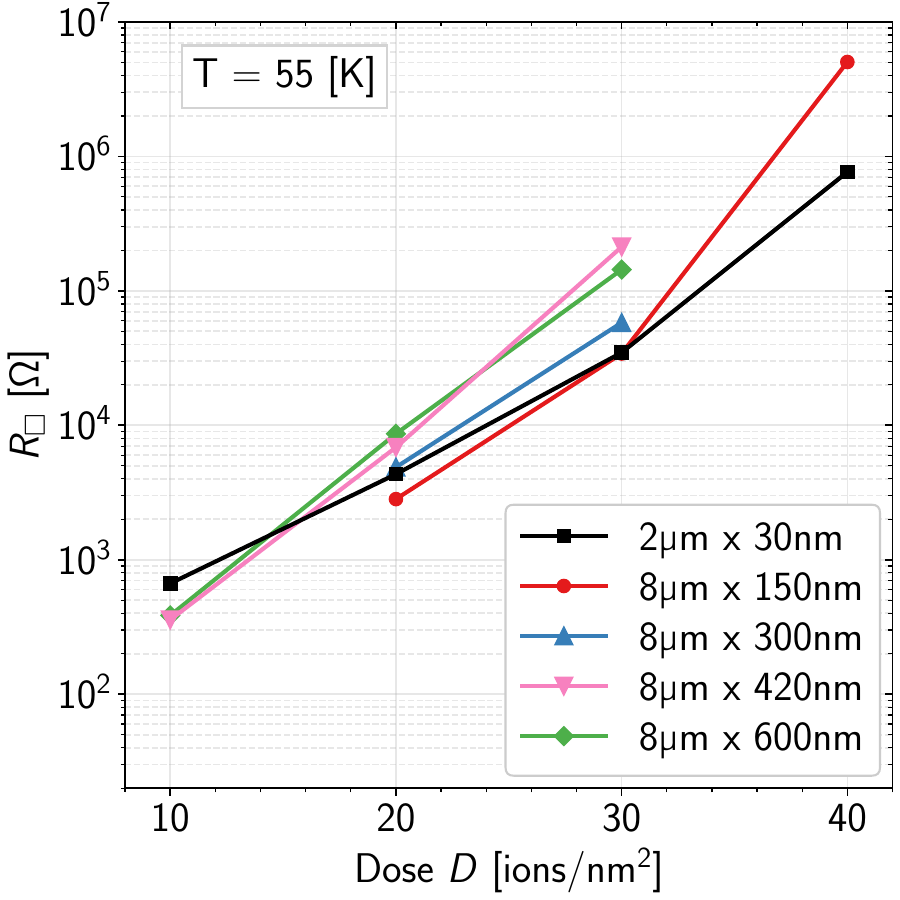}
	\end{center}
	\caption{%
		The dependence of the sheet resistance $R_\square(D)$ for the low $D$ range, where a comparison is possible. Different curves corresponds to different irradiated area size.
	}
	\label{Fig:Rsq}
\end{figure}

To visualize the non-linear growth of resistance with $L/W$, we have extracted the resistance per square (sheet resistance) $R_\square=R\cdot W/L$ at $T=55$~K (chosen to maximise the number of data points), shown in Figure~\ref{Fig:Rsq}. 
In the usual case when $R \propto L/W$, the dependencies $R_\square(D)$ should all overlap and go along one line.
As shown in Figure~\ref{Fig:Rsq}, $R_\square$ also increases strongly (more than one order of magnitude) with $L/W$ at fixed $D$, which is not the case for a usual conductor. 
This, however, is expected if the swelling or bending depends on the size of the irradiated area. 
Note that the dependence corresponding to $W$=\qty{2}{\um} has a less steep increase. This is probably related to the fact that the size effect depends not only on $L$, but also on $W$, and this is the only bridge with $W$=\qty{2}{\um}.
This shows that $R_\square$ depends not only on the size of irradiated area, but also on its shape.

\subsection{Discussion}
\label{sec:discussion}

\begin{figure}
	\begin{center}
		\includegraphics[width=\linewidth]{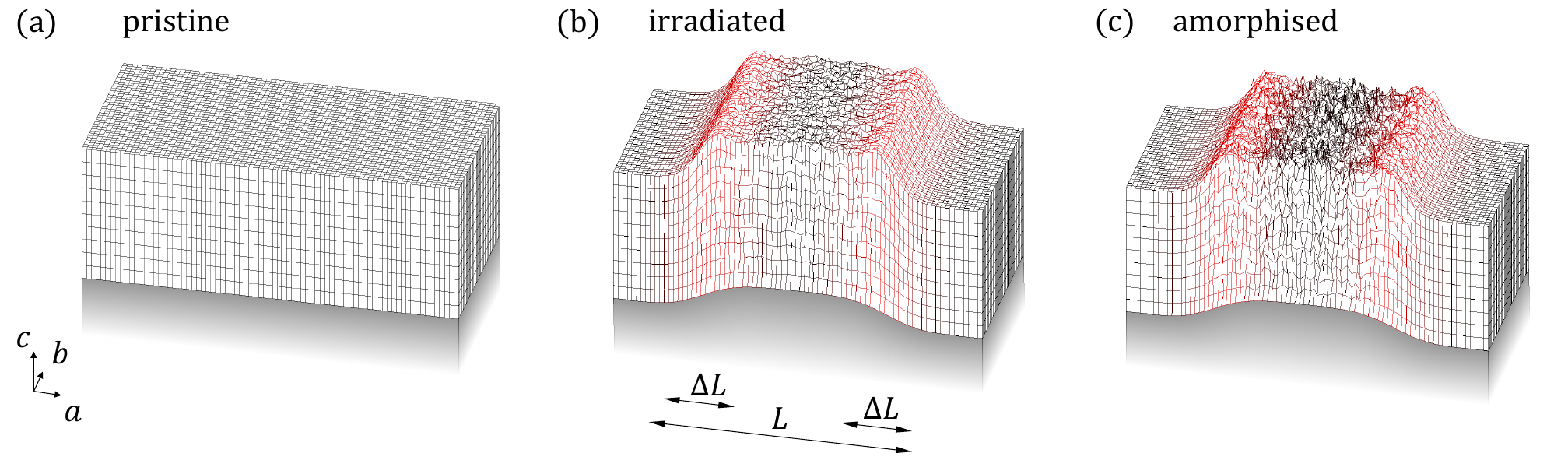}
	\end{center}
	\caption{
		Visual representation of YBCO lattice expansion and substrate bending.
		(a) Pristine YBCO with unchanged lattice constants (grid on top) and pristine LSAT substrate (shaded region below).
		(b) A stripe with length $L$ of YBCO irradiated with $D \le D_c$. Lattice expands in $c$ and $a$ directions by a small amount. The film also experiences bending at the border $\Delta L$ due to He ion bubbles. Areas experiencing the greatest strain are coloured red.
		(c) A stripe of amorphised YBCO with $D > D_c$. Too many defects have been introduced to the lattice causing overwhelming strain. There is no longer a consistent structure and there is significant bending.
	}
	\label{fig:VisualAll}
\end{figure}

Results derived from nanofocused X-ray diffraction suggest that amorphisation is a product of an increasing strain in, and the ultimate "structural collapse" of, the lattice. This is due to the swelling lattice constants and bending in the LSAT substrate, which in turn are due to the large quantities of defects introduced at higher $D$. Figure~\ref{fig:VisualAll} shows an exaggerated representation of how the YBCO lattice changes with increasing dose, from pristine to amorphised.

The critical-disorder limit, $D_{dis}$, is correlated to the length $L$ of irradiated stripes. At small $L$ the lattice is anchored closely by its neighbouring pristine regions. These constraints cause an increased strain on the lattice due to lattice expansion (swelling) and greater film bending, making $D_{dis}$ smaller.
Conversely, at large $L$, the film is less constrained and the lattice has more leeway to expand before the strain becomes too great, so $D_{dis}$ increases.

We hypothesise that this bending allows the in-plane lattice parameters $a$ to expand by up to 6\% (Figure~\ref{fig:ac(D)}b).
From the AFM images of the large irradiated areas (\qtyproduct{5 x 10}{\square\um}), we see that the whole irradiated area is elevated by $\Delta z$ (see Figure~S15  in the SI).
This means that the strongest curvature of the crystal lattice occurs at the border $\Delta L$ between the irradiated and pristine YBCO, while in the centre of the irradiated area the lattice planes are again more or less straight.
Assuming that there is a limit to the lattice plane bending, i.e. there is the maximum possible height change $\Delta z$ which the YBCO film can accommodate, we see that for large irradiated areas the lateral expansion $\delta a$ cannot exceed a certain limit.
Otherwise, the limited bending of the lattice plane cannot compensate the overall increase of the lateral size of the irradiated area, which is proportional to $\propto L\cdot \delta a$.
This argument is supported by the dependence of $\delta a$ on $L$ shown in Figure~\ref{fig:ac(D)}c.

The expansion of the lattice parameter $c$ also causes strain predominantly at the border $\Delta L$ between the irradiated and pristine stripes due to anchoring by the pristine area.
However, the value of this strain does not depend on the size of the irradiated area, since the expansion of the lattice in the vertical direction is unrestricted outside of $\Delta L$, which remains roughly the same regardless of stripe size.
Therefore, we can assume that for large sizes of the irradiated area $L$, the lattice can expand in the vertical direction more freely.
This is also supported by Figure~\ref{fig:ac(D)}c, in which the value $\delta c$ increases with the size $L$.

For the very small irradiated areas (a few tens of nm, $L\lesssim\Delta L$), we expect that there will be only very small changes in the lattice parameters because of a small irradiated area trapped inside the pristine material which does not allow any significant deformations.

\section{Conclusions}
\label{sec:conclusions}

In this work, we used nanofocused X-ray diffraction to measure the change of the YBCO lattice parameters caused by He-FIB as a function of irradiation dose and irradiation area. 
Surprisingly, both the $c$ and $a$ lattice constants increase with $D$. 
The change of the in-plane lattice parameter $a$ is larger than what was observed earlier on the samples irradiated by unfocused He$^+$ beam \cite{Zaluzhnyy:2024:YBCO:He-FIB.StructChg}. This indicates that the structural changes caused by local irradiation, \textit{e.g.} by He-FIB, differ from the changes caused by broad-beam irradiation.

At the critical-disorder dose, $D_{dis}$, one observes a decrease of intensity for both the 004 and 104 peaks corresponding to a sudden loss of order in the large volume fraction of the crystal. We clearly observe a size-effect: $D_{dis}$ increases with the length $L$ of the irradiated stripe embedded into pristine crystal. 
Moreover, this size effect is evident when considering the relative change in the lattice constants, $\delta c$ and $\delta a$. As $L$ increases, we note that $\delta c$ also increases, whereas $\delta a$ decreases.
We also observe a marginally different $D_{dis}$ for the order in $a$ and $c$ directions.

Using AFM, we observe changes in the film's height on a scale approximately one order of magnitude larger than the lattice expansion; which we interpreted as a bending of the substrate. 
This bending agrees with the opposite trends that is seen in $\delta c$ and $\delta a$ when the size of the irradiated area L is changed (see Figure~\ref{fig:ac(D)}c).
The bending in the irradiated stripes with a large length $L$ allows for more expansion in $c$, while keeping the increase in $a$ moderate, due to the strongest curvature occurring exclusively the stripe boundaries. 
By similar logic, the stripes with small $L$ have a larger perimeter-to-area ratio, thus $\delta c$ is smaller, and $\delta a$ larger.

Transport measurements show a superconductor-insulator transition with resistance quickly increasing with the dose $D$. 
In contrast to X-ray diffraction, we do not see any abrupt changes in resistivity when the dose exceeds $D_{dis}$. However, the size effect is also visible: the areas with larger $L$ exhibit up to one order of magnitude larger sheet resistance $R_\square$ than those with small $L$.

Our studies demonstrate that by irradiating narrow stripes one can create a resistor with resistance value tunable by dose and temperature over several orders of magnitude. Making the stripes longer (in transport direction) makes not only the resistance, but also the resistivity larger due to a size-effect related to the swelling and bending of the irradiated region embedded into fixed pristine YBCO crystal.
These findings enable a better understanding of how He$^+$ irradiation parameters can be tuned to modify the properties of superconducting YBCO films

\section{Experimental section}

\subsection{Sample fabrication}

The samples were fabricated by pulsed laser deposition (PLD) of epitaxially-grown $c$-axis oriented YBCO on (100)-oriented single crystal (LaAlO$_3$)$_3$(Sr$_2$AlTaO$_6$)$_7$ (LSAT) substrates.
The thickness of YBCO was 100~nm, which is larger than the typical 30--50~nm used for fabricating Josephson junctions and circuits, due to the need for a stronger X-ray scattering signal.
The sharp diffraction peaks from YBCO on $\theta-2\theta$ scans with a full width at half maximum (FWHM) of $6\cdot10^{-3}$~\AA$^{-1}$ (Figures S1-S2 in the SI) indicate that the films can be considered a high-quality single crystal in the vertical direction.
In the in-plane direction, it is expected to exhibit $\langle 110 \rangle$ twinning, typical for orthorhombic YBCO \cite{Zaluzhnyy:2024:YBCO:He-FIB.StructChg,Khoshnevisan2002}.
The critical temperature for the pristine YBCO film was measured at $T_c\sim88$~K.

\subsection{Irradiation with He$^+$ ions}
The microbridges and stripes on the YBCO film were fabricated by a Zeiss Orion NanoFab helium ion microscope using the 30 keV He$^+$ ion beam.

When irradiating areas, the dose is naturally measured in ions/nm$^2$. It is useful to have a relation between the 1D and 2D dose values that have the same effect, \textit{e.g.}, amorphisation, on the sample. This relation depends on the FIB spot size $\sigma$ (assuming it has a Gaussian shape with the standard deviation $\sigma$, which is typically 2--5~nm). The value of $\sigma$ is supposed to be much larger than the pitch size (pixel size of a helium ion microscope positioning system, which can be made as small as 0.25~nm). In this simplified model the relation between $D^\mathrm{1D}$ given in ions/nm and $D^\mathrm{2D}$ given in ions/nm$^2$ having the same effect on the sample is 
\[
D^\mathrm{1D}=\sqrt{2\pi}\sigma\cdot D^\mathrm{2D}
\]
For example, if the amorphisation of lines starts at $D_c^\mathrm{1D}\approx1800$~ions/nm then the amorphisation of areas will start at $D_c^\mathrm{1D}\approx180$~ions/nm$^2$, assuming $\sigma\approx4$~nm.

\subsection{Synchrotron experiment}

The nanofocused X-ray diffraction experiment \cite{Schuelli2018} was carried out at the beamline ID-13 of the European Synchrotron radiation facility (ESRF).
The X-ray beam with energy $E=15$~keV was focused to the size of approximately \qtyproduct{100 x 100}{\square\nm} (FWHM) using multilayer Laue lenses \cite{Kubec2017, Kubec2018}. 
The stripes on the YBCO chip fabricated by He-FIB were raster scanned on a rectangular grid with resolution \qtyproduct{100 x 1000}{\square\nm} (vert. $\times$ hor.), owing an increased footprint of the X-ray beam in horizontal direction when the sample is aligned for a Bragg condition.
Thus, the spatial resolution of approximately \qty{100}{\nm} was achieved in the direction perpendicular to the stripes (Figure~S2 in the SI).
The large EigerX~4M detector placed \qty{260}{\mm} behind the sample was used to record the diffraction patterns from the film.

\subsection{Atomic force microscopy}
The AFM measurements were performed with the Nanowizard-II system from JPK-Bruker.

\subsection{Transport measurement}
The electrical transport measurements were performed at the University of T\"ubingen after the XRD experiment at the ESRF.
For the transport measurements we used microbridges fabricated on the same chip that was used for the XRD  experiment.
While cooling from $T\approx$\qtyrange{300}{6}{\K} (and warming up again) in a liquid He dewar, each microbridge was biased by a constant probe current of $I_\mathrm{b}$=\qty{1}{\micro\A}; the voltage across each microbridge (between bonding pads) was measured in 4-point-configuration, thus determining the resistance.
During each cooling cycle, four microbridges were measured simultaneously.
Note that for $T<T_\mathrm{c} \approx 88$~K the pristine (non-irradiated) YBCO film is superconducting; therefore, all measured resistance corresponds to the resistance of irradiated areas. However, at $T>T_\mathrm{c}$, the non-superconducting pristine YBCO (as well as the Au covered leads) also contribute to the total resistance shown in Figure~\ref{Fig:R(T)2x30} (few k$\Omega$ for YBCO, and few $\Omega$ for Au).

\medskip
\medskip
\medskip
\textbf{Supporting information} \par 
Supporting Information is available.

\medskip
\textbf{Data availability}
The data collected during an experiment at the ID13 beamline of European Synchrotron Radiation Source ESRF (Grenoble, France) are available at https://doi.esrf.fr/10.15151/ESRF-ES-1691831423 after an embargo period \cite{dataESRF}.

\medskip
\textbf{Author contributions}

R.K., D.K., E.G. and I.A.Z. conceptualized the work;
R.H., S.K., C.S. prepared the samples;
R.C., R.H., P.Z., J.U. M.B. and I.A.Z. performed measurements;
R.C., R.H., A.A., E.G. and I.A.Z. analyzed the data;
R.C., R.H. and E.G. wrote the manuscript with contribution from all authors.

\medskip
\textbf{Acknowledgements} \par 
We acknowledge the European Synchrotron Radiation Facility (ESRF) for provision of synchrotron radiation facilities under proposal number HC-5713.
We acknowledge the LISA+ Center at the University of T\"ubingen for technical support.
We thank DAPHNE4NFDI for funding.
E.G. thanks DFG (project Go-1106/7) for financial support. We thank Christopher Buckreus for help with processing $R(T)$ curves and Frank Schreiber for the support of this work.

\medskip
\textbf{Conflicts of interest} \par
The authors declare no conflict of interest.

\medskip

\bibliographystyle{MSP}
\bibliography{YBCO_LocalXRD.bib}

\begin{thebibliography}{10}
\providecommand{\url}[1]{\texttt{#1}}
\providecommand{\urlprefix}{URL }

\bibitem{Cybart:2015:He-FIB.JJs}
S.~A. Cybart, E.~Y. Cho, T.~J. Wong, B.~H. Wehlin, M.~K. Ma, C.~Huynh, R.~C.
  Dynes,
\newblock \emph{Nat. Nanotechnol.} \textbf{2015}, \emph{10} 598.

\bibitem{Weber:2025:SQUID-on-lever}
T.~Weber, D.~Jetter, J.~Ullmann, S.~A. Koch, S.~F. Pfander, K.~Kress,
  A.~Vervelaki, B.~Gross, O.~Kieler, U.~Drechsler, P.~R. Baral, A.~Magrez,
  R.~Kleiner, A.~W. Knoll, M.~Poggio, D.~Koelle,
\newblock \emph{Phys. Rev. Appl.} \textbf{2025}, \emph{24} 054041.

\bibitem{Kasaei:2018:He-FIB:MgB2-JJ}
L.~Kasaei, T.~Melbourne, V.~Manichev, L.~C. Feldman, T.~Gustafsson, K.~Chen,
  X.~X. Xi, B.~A. Davidson,
\newblock \emph{AIP Adv.} \textbf{2018}, \emph{8}, 7 075020.

\bibitem{Kasaei:2019:bJJA.LowSpread}
L.~Kasaei, T.~Melbourne, M.~Li, V.~Manichev, F.~Qin, H.~Hijazi, L.~C. Feldman,
  T.~Gustafsson, B.~A. Davidson, X.~Xi, K.~Chen,
\newblock \emph{IEEE Trans. Appl. Supercond.} \textbf{2019}, \emph{29}, 5 1.

\bibitem{Yin:2024:He-FIB:MgB2-JJ}
D.~Yin, X.~Cai, T.~Xu, R.~Sun, Z.~Chen, Y.~Han, L.~Tian, Y.~Wang, Y.~Zhang,
  Z.~Gan,
\newblock \emph{Phys. C} \textbf{2024}, \emph{623} 1354532.

\bibitem{Chen:2024:bJJs(Co-BaFeS)}
Z.~Chen, Y.~Zhang, P.~Ma, Z.~Xu, Y.~Li, Y.~Wang, J.~Lu, Y.~Ma, Z.~Gan,
\newblock \emph{Chin. Phys. B} \textbf{2024}, \emph{33}, 4 047405.

\bibitem{Ruhtinas:2025:He-FIB:NbTiN:bJJs}
A.~Ruhtinas, I.~J. Maasilta,
\newblock \emph{Phys. Rev. Res.} \textbf{2025}, \emph{7} 043252.

\bibitem{Karrer:2024:He-FIB:PinArr20nm}
M.~Karrer, B.~Aichner, K.~Wurster, C.~Mag\'en, C.~Schmid, R.~Hutt,
  B.~Budinsk\'a, O.~V. Dobrovolskiy, R.~Kleiner, W.~Lang, E.~Goldobin,
  D.~Koelle,
\newblock \emph{Phys. Rev. Appl.} \textbf{2024}, \emph{22} 014043.

\bibitem{Aichner:2023:MagRes(ang)}
B.~Aichner, L.~Backmeister, M.~Karrer, K.~Wurster, R.~Kleiner, E.~Goldobin,
  D.~Koelle, W.~Lang,
\newblock \emph{Condensed Matter} \textbf{2023}, \emph{8}, 2 32.

\bibitem{Backmeister:2022:He-FIB:YBCO:BoseGlass}
L.~Backmeister, B.~Aichner, M.~Karrer, K.~Wurster, R.~Kleiner, E.~Goldobin,
  D.~Koelle, W.~Lang,
\newblock \emph{Nanomaterials} \textbf{2022}, \emph{12}, 19.

\bibitem{Cho:2018:He-FIB:bJJs(w)}
E.~Y. Cho, Y.~W. Zhou, J.~Y. Cho, S.~A. Cybart,
\newblock \emph{Appl. Phys. Lett.} \textbf{2018}, \emph{113}, 2 022604.

\bibitem{Mueller:2019:He-FIB:JJ&SQUID}
B.~M\"uller, M.~Karrer, F.~Limberger, M.~Becker, B.~Schr\"oppel, C.~Burkhardt,
  R.~Kleiner, E.~Goldobin, D.~Koelle,
\newblock \emph{Phys. Rev. Appl.} \textbf{2019}, \emph{11} 044082.

\bibitem{Couedo:2020:He-FIB:YBCO-JJs}
F.~Couëdo, P.~Amari, C.~Feuillet-Palma, C.~Ulysse, Y.~K. Srivastava, R.~Singh,
  N.~Bergeal, J.~Lesueur,
\newblock \emph{Sci. Rep.} \textbf{2020}, \emph{10}, 1 10256.

\bibitem{Chen:2022:YBCO-bJJs}
Z.~Chen, Y.~Li, R.~Zhu, J.~Xu, T.~Xu, D.~Yin, X.~Cai, Y.~Wang, J.~Lu, Y.~Zhang,
  P.~Ma,
\newblock \emph{Chin. Phys. Lett.} \textbf{2022}, \emph{39}, 7 077402.

\bibitem{Cho:2015:He-FIB.SQUID}
E.~Y. Cho, M.~K. Ma, C.~Huynh, K.~Pratt, D.~N. Paulson, V.~N. Glyantsev, R.~C.
  Dynes, S.~A. Cybart,
\newblock \emph{Appl. Phys. Lett.} \textbf{2015}, \emph{106}, 25 252601.

\bibitem{Proepper:2024:bJJ.THz}
M.~Pr\"opper, D.~Hanisch, C.~Schmid, P.~J. Ritter, M.~Neumann, E.~Goldobin,
  D.~Koelle, R.~Kleiner, M.~Schilling, B.~Hampel,
\newblock \emph{IEEE Trans. Appl. Supercond.} \textbf{2024}, \emph{34}, 3
  1100505.

\bibitem{Proepper:2024:He-FIB:YBCO:JJA}
M.~Pr\"opper, D.~Hanisch, C.~Schmid, M.~Neumann, P.~J. Ritter, M.-A. Tucholke,
  E.~Goldobin, D.~Koelle, R.~Kleiner, M.~Schilling, B.~Hampel,
\newblock \emph{IEEE Trans. Appl. Supercond.} \textbf{2025}, \emph{35}, 5
  1100105.

\bibitem{Schmid:2025:He-FIB:YBCO-JJD}
C.~Schmid, A.~Jozani, R.~Kleiner, D.~Koelle, E.~Goldobin,
\newblock \emph{Phys. Rev. Appl.} \textbf{2025}, \emph{24} 014041.

\bibitem{Schmid:He-FIB:YBCO:cJJ&SQUID}
C.~Schmid, C.~Buckreus, D.~Haas, M.~Pröpper, R.~Hutt, C.~Magén, D.~Hanisch,
  M.~Karrer, M.~Schilling, D.~Koelle, R.~Kleiner, E.~Goldobin,
\newblock {YBa$_2$Cu$_3$O$_{7}$} nano-constriction {J}osephson junctions and
  {SQUIDs} fabricated by focused helium-ion-beam irradiation, \textbf{2025},
\newblock \urlprefix\url{https://arxiv.org/abs/2511.19197}.

\bibitem{Karrer:2024:bJJ(t)}
M.~Karrer, K.~Wurster, J.~Linek, M.~Meichsner, R.~Kleiner, E.~Goldobin,
  D.~Koelle,
\newblock \emph{Phys. Rev. Appl.} \textbf{2024}, \emph{21} 014065.

\bibitem{Valles1989}
J.~M. Valles, A.~E. White, K.~T. Short, R.~C. Dynes, J.~P. Garno, A.~F.~J.
  Levi, M.~Anzlowar, K.~Baldwin,
\newblock \emph{Phys. Rev. B} \textbf{1989}, \emph{39} 11599.

\bibitem{Gupta1992}
R.~P. Gupta, M.~Gupta,
\newblock \emph{Phys. Rev. B} \textbf{1992}, \emph{45} 9958.

\bibitem{Menushenkov1995a}
A.~Menushenkov, A.~Ignatov, A.~Ivanov, D.~Kochubey, V.~Chernov, S.~Nikitenko,
\newblock \emph{Nucl. Instrum. Methods Phys. Res. A} \textbf{1995}, \emph{359},
  1-2 236.

\bibitem{Navacerrada2000}
M.~A. Navacerrada, D.~Arias, Z.~Sefrioui, G.~Loos, M.~L. Luc\'{i}a,
  J.~Santamar\'{i}a, F.~S\'{a}nchez-Quesada, M.~Varela,
\newblock \emph{Appl. Phys. Lett.} \textbf{2000}, \emph{76}, 22 3289.

\bibitem{Nicholls2022}
R.~J. Nicholls, S.~Diaz-Moreno, W.~Iliffe, Y.~Linden, T.~Mousavi, M.~Aramini,
  M.~Danaie, C.~R.~M. Grovenor, S.~C. Speller,
\newblock \emph{Commun. Mat.} \textbf{2022}, \emph{3}, 1 52.

\bibitem{Gray2022}
R.~L. Gray, M.~J.~D. Rushton, S.~T. Murphy,
\newblock \emph{Supercond. Sci. Technol.} \textbf{2022}, \emph{35}, 3 035010.

\bibitem{Lee2016}
Y.~J. Lee, J.~H. Choi, B.-H. Jun, J.~Joo, C.~S. Kim, C.-J. Kim,
\newblock \emph{Prog. Supercond. Cryog.} \textbf{2016}, \emph{18}, 4 15.

\bibitem{Suvorova2014}
E.~I. Suvorova, M.~Cantoni, P.~A. Buffat, A.~Y. Didyk, L.~K. Antonova, A.~V.
  Troitskii, G.~N. Mikhailova,
\newblock \emph{Acta Mater.} \textbf{2014}, \emph{75} 71–79.

\bibitem{Kirk1999}
M.~Kirk, Y.~Yan,
\newblock \emph{Micron.} \textbf{1999}, \emph{30}, 5 507.

\bibitem{Hutt:AmorDose}
R.~Hutt, et~al.,
\newblock Amorhization of {YBa$_2$Cu$_3$O$_{7}$} due to {He-FIB} irradiation,
\newblock unpublished.

\bibitem{Zaluzhnyy:2024:YBCO:He-FIB.StructChg}
I.~A. Zaluzhnyy, U.~Goteti, B.~K. Stoychev, R.~Basak, E.~S. Lamb, E.~Kisiel,
  T.~Zhou, Z.~Cai, M.~V. Holt, J.~W. Beeman, E.~Y. Cho, S.~Cybart, O.~G.
  Shpyrko, R.~Dynes, A.~Frano,
\newblock \emph{ACS Appl. Nano Mater.} \textbf{2024}, \emph{7}, 14 15943.

\bibitem{Terai1991}
T.~Terai, K.~Kusagaya, T.~Furuta, Y.~Takahashi, Y.~Enomoto, S.~Kubo,
\newblock \emph{Physica C} \textbf{1991}, \emph{185-189} 2473.

\bibitem{Zhao1991}
Y.~Zhao, W.~Chu, M.~Davis, J.~Wolfe, S.~Deshmukh, D.~Economou, A.~Mcguire,
\newblock \emph{Physica C} \textbf{1991}, \emph{184}, 1-3 144.

\bibitem{Meyer1989}
O.~Meyer, B.~Egner, G.~Xiong, X.~Xi, G.~Linker, J.~Geerk,
\newblock \emph{Nucl. Instrum. Methods Phys. Res. B} \textbf{1989}, \emph{39},
  1-4 628.

\bibitem{Arias2003}
D.~Arias, Z.~Sefrioui, G.~D. Loos, F.~Agullo-Rueda, J.~Garcia-Barriocanal,
  C.~Leon, J.~Santamaria,
\newblock \emph{Phys. Rev. B} \textbf{2003}, \emph{68} 094515.

\bibitem{Mletschnig2019}
K.~Mletschnig, W.~Lang,
\newblock \emph{Microelectron. Eng.} \textbf{2019}, \emph{215} 110982.

\bibitem{Li2024}
Z.~Li, H.~Ma, J.~Huang, Y.~Liu, M.~Shao, Z.~Luo, X.~Du, C.~Wu, N.~Li, H.~Wang,
  P.~Li,
\newblock \emph{Supercond. Sci. Technol.} \textbf{2024}, \emph{37}, 12 125010.

\bibitem{Abukaev2025}
A.~Abukaev, C.~V\"{o}lter, M.~Romodin, S.~Schwartzkopff, F.~Bertram,
  O.~Konovalov, A.~Hinderhofer, D.~Lapkin, F.~Schreiber,
\newblock \emph{J. Appl. Crystallogr.} \textbf{2026}, \emph{59}, 1 263.

\bibitem{Jorgensen1990}
J.~D. Jorgensen, B.~W. Veal, A.~P. Paulikas, L.~J. Nowicki, G.~W. Crabtree,
  H.~Claus, W.~K. Kwok,
\newblock \emph{Phys. Rev. B} \textbf{1990}, \emph{41} 1863.

\bibitem{Cava1990}
R.~Cava, A.~Hewat, E.~Hewat, B.~Batlogg, M.~Marezio, K.~Rabe, J.~Krajewski,
  W.~Peck, L.~Rupp,
\newblock \emph{Physica C} \textbf{1990}, \emph{165}, 5 419.

\bibitem{Khoshnevisan2002}
B.~Khoshnevisan, D.~K. Ross, D.~P. Broom, M.~Babaeipour,
\newblock \emph{J. Condens. Matter Phys.} \textbf{2002}, \emph{14}, 41 9763.

\bibitem{Lesueur:1993:YBCO:SIT(He-ions)}
J.~Lesueur, L.~Dumoulin, S.~Quillet, J.~Radcliffe,
\newblock \emph{J. Alloys Compd.} \textbf{1993}, \emph{195} 527.

\bibitem{Schuelli2018}
T.~U. Sch\"{u}lli, S.~J. Leake,
\newblock \emph{Curr. Opin. Solid State Mater. Sci.} \textbf{2018}, \emph{22},
  5 188–201.

\bibitem{Kubec2017}
A.~Kubec, K.~Melzer, J.~Gluch, S.~Niese, S.~Braun, J.~Patommel, M.~Burghammer,
  A.~Leson,
\newblock \emph{J. Synchrotron Radiat.} \textbf{2017}, \emph{24}, 2 413–421.

\bibitem{Kubec2018}
A.~Kubec, S.~Niese, M.~Rosenthal, J.~Gluch, M.~Burghammer, P.~Gawlitza,
  J.~Keckes, A.~Leson,
\newblock \emph{J. Instrum.} \textbf{2018}, \emph{13}, 04 C04011–C04011.

\bibitem{dataESRF}
R.~Hutt, J.~Ullmann, I.~Zaluzhnyy, P.~Zimmermann,
\newblock Structural changes in {YBa$_2$Cu$_3$O$_7$} films irradiated with
  focused {He}$^+$ ion beam [{D}ataset]. {E}uropean {S}ynchrotron {R}adiation
  {F}acility, \textbf{2027}.

\end{thebibliography}


\newpage
\begin{figure}
\textbf{Table of Contents}\\
\medskip
\begin{center}
  \includegraphics[width=0.5\linewidth]{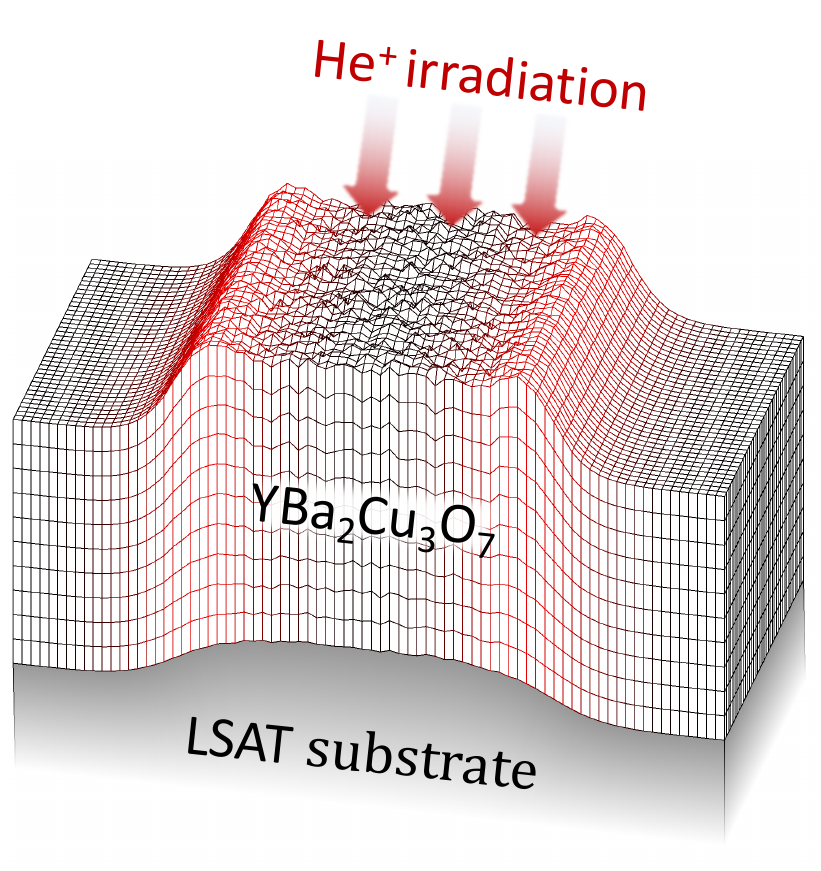}
  \end{center}
  \medskip
  \caption*{Irradiation with a focused helium ion beam turns superconducting {YBa}$_2${Cu}$_3${O}$_7$ into an insulator. Using nanofocused X-ray beam, we studied the structural changes in {YBa}$_2${Cu}$_3${O}$_7$ caused by the ion beam and investigated the effects of the irradiation dose and irradiated area.
  We observed expansion of the unit cell, partial amorphisation and bending of the {YBa}$_2${Cu}$_3${O}$_7$ film.
  These effects are quantitatively and qualitatively different from irradiation with unfocused ion beams.}
\end{figure}

\end{document}